\documentclass[fleqn,usenatbib]{mnras}%https://ja.overleaf.com/project/6226f92c00547e19b44639f2

\usepackage{newtxtext,newtxmath}
\usepackage{threeparttable}

\usepackage[T1]{fontenc}
\usepackage{ae,aecompl}

\usepackage{graphicx}	% Including figure files
\usepackage{amsmath}	% Advanced maths commands
\usepackage{lscape}
\usepackage{longtable}
\usepackage{multirow}

\usepackage{academicons}
\usepackage{orcidlink}
\usepackage{lineno}

\title[X-ray variability of Mrk 766]{Origin of the complex iron line structure and spectral variation  in  Mrk 766} 

\author[Y. Mochizuki et al.]{Yuto Mochizuki$^{\orcidlink{0000-0003-3224-1821}}$ $^{1,2}$
\thanks{E-mail: mochizuki@ac.jaxa.jp},
Misaki Mizumoto$^{\orcidlink{0000-0003-2161-0361}}$ $^3$, and
Ken Ebisawa$^{\orcidlink{0000-0002-5352-7178}}$ $^{1,2}$
\\
$^{1}$Institute of Space and Astronautical Science (ISAS), Japan Aerospace Exploration Agency (JAXA), 3-1-1 Yoshinodai, Chuo-ku, Sagamihara, Kanagawa \\
252-5210, Japan\\
$^{2}$Department of Astronomy, Graduate School of Science, The University of Tokyo, 7-3-1 Hongo, Bunkyo-ku, Tokyo 113-0033, Japan\\
$^{3}$Science education research unit,  University of Teacher Education Fukuoka, 1-1 Akama-bunkyo-machi, Munakata, Fukuoka 811-4192, Japan
}

\date{Accepted 2023 July 27. Received 2023 July 10; in original form 2023 April 3}

\pubyear{2023}

\begin{document}
\label{firstpage}
\pagerange{\pageref{firstpage}--\pageref{lastpage}}
\maketitle

\begin{abstract}
Complex Fe-K emission/absorption line features are commonly observed in the 6--11 keV band from Active Galactic Nuclei (AGN). These features are formed in various physical components surrounding the black holes.
The Narrow-Line Seyfert 1 (NLS1) galaxy Mrk 766, in particular, exhibits characteristic
blue-shifted Fe-K absorption lines caused by the ultra-fast outflow (UFO),  and 
 a broad Fe-K emission line, as well as  variable  absorbers partially covering the X-ray emitting region.
 We re-analyze the Mrk 766 archival data of \textit{XMM-Newton}, \textit{NuSTAR}, and \textit{Swift} to investigate the origin of the Fe-K line feature and the 0.3--79 keV energy spectral variation.
We have found that the spectral variation in $\lesssim$10 keV is primarily 
explained by  the variable partial covering of the central X-ray source by multi-layer absorbing clouds. The Fe-K line feature
consists of the blue-shifted absorption lines due to the UFO, a narrow emission line from the distant material,  a 
broad emission line from the inner-disk reflection, and a 
slightly broadened weak emission line at around 6.4--6.7 keV whose  equivalent width is 
$\sim$0.05 keV. 
The last one  is  presumably
due to the
resonance scattering in the UFO out of the line-of-sight, 
as  predicted  by a  Monte Carlo simulation based on  the hydrodynamical UFO modeling.  
We suggest that the seemingly complex Fe-K line features and the 
X-ray energy spectra of  Mrk 766 are explained by 
a moderately extended central X-ray source around a Schwarzschild black hole, an optically thick accretion disk with a 
truncated inner-radius, 
the UFO, multi-layer partial covering clouds, and a torus.
\end{abstract}


%%%%%%%%%%%%%%%%%%%%%%%%%%%%%%%%%%%%%%%%%%%%%%%%%%

%%%%%%%%%%%%%%%%% BODY OF PAPER %%%%%%%%%%%%%%%%%%

\section{Introduction}

X-ray radiation from Active Galactic Nuclei (AGN) is reprocessed by surrounding materials around the central black hole, such as 
the accretion disk, torus, outflow, and various ionized gases, which affect the 
X-ray energy spectrum.  
Absorption lines are produced by the intervening hot plasma, while the emission lines result from fluorescence or recombination in the surrounding materials. 
In particular, 
complex Fe-K spectral features, often observed in the 6-9 keV band,  arise due to
multiple absorbers and scatterers around the black holes.
By resolving the complex Fe-K line structure, we can gain a deeper understanding of the physical properties and processes in the vicinity of the black hole.

Mrk 766 is classified as a Narrow-line Seyfert 1 (NLS1) galaxy with 
a black hole mass of  $ \log (M_\mathrm{BH}/M_\odot ) = 6.82^{+0.05}_{-0.06} $ \citep{ben15}
at a cosmic redshift of $0.01271 \pm 0.00005$ \citep{adel11}.
Over the past few decades, this source has been extensively observed with  \textit{XMM-Newton}, \textit{NuSTAR}, and \textit{Swift}.

Mrk 766 has the following two notable X-ray characteristics: 
First, significant X-ray flux/spectral variability with a timescale of less than a day has been observed,
such that the soft ($\lesssim$5 keV) X-ray band appears to be more variable than the hard band \citep[e.g., ][]{ris11}.
This spectral variation is commonly explained by  the partial covering of the central
X-ray source by clumpy absorbers (e.g., \citealt{risa-long09}; \citealt{risa-vari09}). 
Various models have been proposed to describe  the geometry of these clouds;   double ionized layers \citep{Miller07, Tur07}, cometary shape with  ionized and neutral layers (\citealt{ris11}, \citealt{mai10}), a single ionized  \citep{lie14} or  neutral layer \citep{bui18}.

Second, the X-ray spectra show characteristic Fe-K emission and absorption structures in 6--11 keV.
A hint of the ionized Fe absorption lines has been reported \citep{pou03}, and
recent studies indicate presence of the blueshifted ionized Fe-K absorption lines  caused by the outflowing ionized absorbers in the line-of-sight \citep[e.g.,][]{Miller07, Tur07, ris11, lie14, bui18}.
The outflow velocity,  $\sim$$0.1 c$, is categorized as an ultra-fast outflow \citep[UFO;][]{pou03,king03,rev09}.
The ionization parameter of these absorbing clouds is $ \xi \sim 10^{3-5}$ erg$\cdot$cm s$^{-1}$ and the 
hydrogen column density is $N_{\rm{H}} \sim 10^{22-24}$ cm$^{-2}$ (e.g., \citealt{tom11}).
In addition, a broad Fe-K emission line is observed in 6--7 keV.
\citet{lie14} and \citet{bui18} explained this broad line assuming the disk reflection caused by a tiny corona in the very vicinity of the fast-rotating black hole, subject to the strong relativistic distortion (e.g.,  \citealt{fa09}).

Both the partial covering and the Fe-band spectral features have to be reconciled in the 
X-ray spectral modeling of Mrk 766.  However, the relationship between the partial covering clouds, UFOs, and the broad Fe emission lines remains unclear.
In particular, the relativistic line-broadening requires the tiny X-ray emission region located in the close vicinity of the black hole.
Such a very compact X-ray source should not be {\em  partially} but {\em fully}\/ covered by the absorbing clouds in the line-of-sight (e.g., \citealt{iso16}).
Thus, it is required to scrutinize the origin of the 
partial covering clouds, UFOs, and the broad Fe emission lines to build  a  consistent picture of the X-ray emission mechanism in Mrk 766
and other similar Seyfert galaxies. 

This paper aims to reveal the origin of the complex Fe-K line structure in Mrk 766, and to propose a consistent 
X-ray spectral model based on the realistic AGN spectral components.
Section 2 describes the method of data analysis,  Section 3 presents our spectral fitting models and the fitting results. The results are discussed in Section 4 and the paper is concluded in Section 5.

\section{Observation and Data reduction}

We used the archival data of \textit{XMM-Newton}, \textit{NuSTAR}, and \textit{Swift}. The datasets are listed in Table \ref{tab:xmm}.

All the available  \textit{XMM-Newton} \citep{xmm} archival data to date were used.
In the data analysis, 
the Science Analysis Software ({\tt SAS}, v.19.1.0) \citep{sas} and the calibration files corresponding to this SAS version were used.
In the European Photon Imaging Camera (EPIC)-pn \citep{epic},
{\tt epchain} 
was used to produce the clean events.
From each observation ID, a single spectrum was produced; in total, we made nine spectral datasets.
The source spectra were extracted from a circular region of a radius of $30^{\prime\prime}$ centered on the source with {\tt PATTERN$<=$4}.
The background spectra were extracted from a circular region of a radius of $45^{\prime\prime}$ in the same CCD chip near the source, 
avoiding the chip edges and/or the serendipitous sources,  and minimizing the effects of the Cu-K background  lines.
High background periods were excluded when the count rates in 10--12~keV with {\tt PATTERN==0} were higher than 0.4 cts\,s$^{-1}$.
By using  {\tt epatplot}, a possible pile-up effect in the source region was examined.
The only full-window mode data,  XMM9,  was affected by the pile-up and 
hollowed at the center with a circle of $10^{\prime\prime}$ radius. Other observations were performed in the small-window mode and were not affected by the pile-up.
In addition, we made a merged spectrum by combining all the  XMM-pn spectra with {\tt epicspeccombine}.
In the Reflection Grating Spectrometer (RGS) \citep{rgs}, {\tt rgsproc}  was used to create the 
cleaned events with 
the same good time intervals of XMM-pn. 
The RGS1 and the RGS2 spectra of each observation ID were combined using  {\tt rgscombine}.

We also used all the available \textit{NuSTAR} \citep{NuSTAR} archival data. 
The \textit{NuSTAR} data analysis software (v.2.1.2) in  HEASOFT (v.6.30.1) with  the 
contemporary calibration files (20220926) were used. {\tt nupipeline} was run to create the clean events for both the focal plane modules (FPMA and FPMB).
The source spectra were extracted from a circular region of  $40^{\prime\prime}$ radius centered on the source by using  {\tt nuproduct}.
The background spectra were extracted from a circular region of  $150^{\prime\prime}$ radius near the source region.

We also used the \textit{Swift} \citep{Swift} archival data
which was taken simultaneously  with \textit{NuSTAR} (Nu10).
The \textit{Swift} XRT data was reduced using HEASOFT (v.6.30.1) to produce the clean events.
The source spectrum was extracted from a circular region of a radius of $40^{\prime\prime}$ centered on the source.
The background spectrum was extracted from a circular region of a radius of $150^{\prime\prime}$ from the nearby blank region.

The XMM-pn, \textit{NuSTAR} and \textit{Swift} spectra were rebinned to enhance the visibility around the UFO feature
depending on the counting rates and  exposure time; 
at least 400 counts per new energy-bin for  XMM2 and Nu10, 50 counts for  XMM9 and Nu11 and Sw12, and 200 counts for the other spectra.
All the  XMM-RGS spectra were binned to have at least 25 counts per new energy-bin.

\begin{table*}
	\centering
	\caption{List of the archive data used in this paper}
	\label{tab:xmm}
\begin{threeparttable}
	\begin{tabular}{ccccc} 
		\hline
\multicolumn{1}{c}{Name}	&	ID & Date&Exposure (s)$^{*1}$ & Count rate (s$^{-1}$)$^{*2}$\\
		\hline
XMM1			&0096020101 	&2000.05.20 &25,544 &13.3 \\
XMM2            &0109141301	    &2001.05.20 &88,243 &22.6\\
XMM3            &0304030101 	&2005.05.23 &65,193 &3.9\\
XMM4 			&0304030301 	&2005.05.25 &68,938 &8.6 \\
XMM5 			&0304030401 	&2005.05.27 &65,767 &11.3 \\
XMM6 			&0304030501	 	&2005.05.29 &64,526 &14.0 \\
XMM7 			&0304030601 	&2005.05.31 &60,698 &11.7 \\
XMM8 			&0304030701 	&2005.06.03 &20,089 &10.0 \\
XMM9$^{*3}$     &0763790401     &2015.07.05 &19,534 &4.7\\\hline
Nu10            &60001048002    &2015.01.24 &90,107 &0.4\\
Nu11            &60101022002    &2015.07.05 &23,545 &0.3\\\hline
Sw12            &00080076002    &2015.01.25 &4,885  &1.0\\
\hline
	\end{tabular}
\begin{tablenotes}
\item[*1] After removing the  background flare periods.
\item[*2] In 0.3--11 keV for XMM-pn data, and in 3--79 keV for \textit{NuSTAR} data (averaging FPAM and FPMB). 
\item[*3] Full-window mode.
\end{tablenotes}
\end{threeparttable}
\end{table*}

In the following, we tried to construct a model which explains the 
XMM-pn and XMM-RGS spectra simultaneously, as well as the \textit{NuSTAR} and \textit{Swift} spectra when available.
The energy ranges we used for XMM-pn, XMM-RGS, \textit{NuSTAR}, \textit{Swift} are 0.3--11 keV, 0.33--2 keV, and 
3--79 keV, 0.3--7.5 keV, respectively.
All the spectral fittings were performed by using {\tt xspec} v.12.12.1 \citep{xspec},
where we employed the $\chi^2$-statistics.
 The error values correspond to the 90 \% confidence level.

\section{Results}

\subsection{Complex Fe-K line strucuture } \label{sec:3.1}

\begin{figure}
\centering
	\includegraphics[width=8cm]{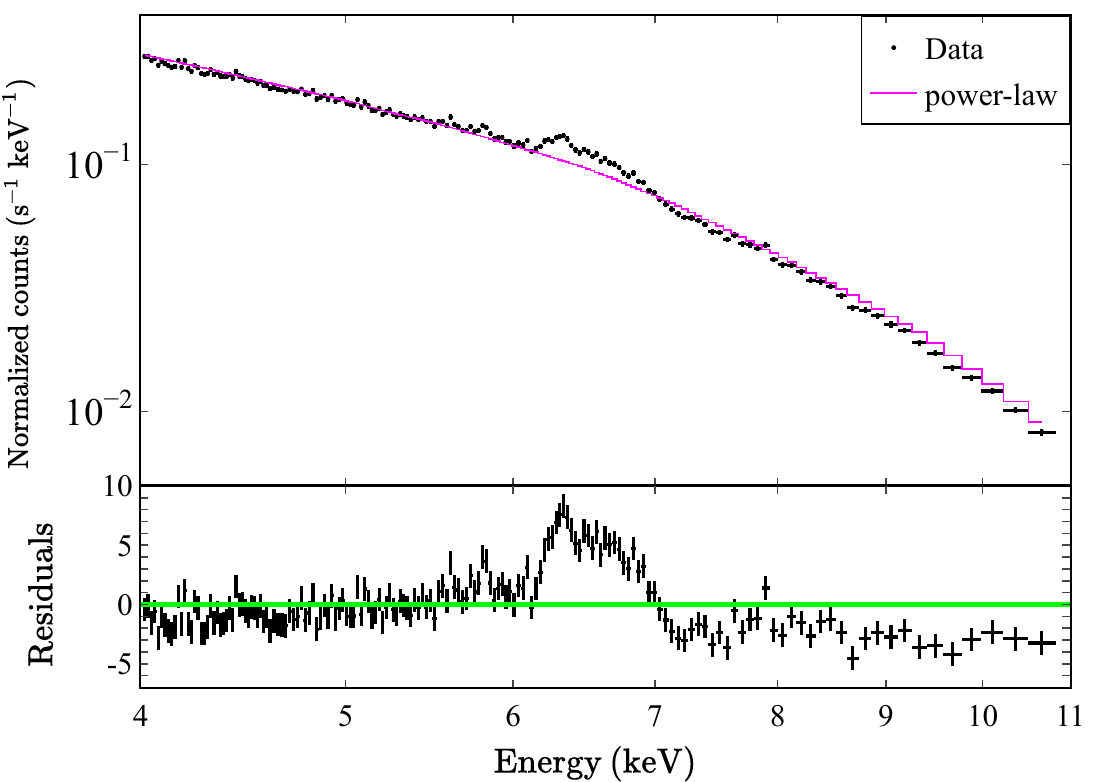}
    \caption{Spectral fitting result of the merged XMM-pn spectrum with a power-law. 
    The spectrum was re-binned to have 1,500 counts per new spectral bin. 
    In the top panel,  the merged  XMM-pn spectrum is shown in black and  {\tt powerlaw} function
     in magenta. The bottom panel shows the residuals in the unit of  
     standard deviation.
    }
    \label{fig:allmerge} 
\end{figure}

First, we demonstrate the complex Fe line structure in Mrk 766.  We analyzed the merged spectrum of all the XMM-pn data in  4--11~keV by fitting with
a power-law  (Fig.\ \ref{fig:allmerge}).
We can clearly see the complex Fe-K structure, 
a narrow emission line at 6.4~keV, a broad emission feature in 6--7~keV, and blue-shifted absorption lines in 7--11~keV.
Also, we can recognize a weak emission line at $\sim$5.8 keV which is presumably the Mn-K line.
Below, we are going to explain these spectral features as well as the continuum spectral components.

\subsection{A double-layer partial covering model}
We adopted  a double-layer partial covering model, which is known to explain the broadband X-ray spectral variability (e.g., \citealt{ris11}).
Since the change in the column density of an optically-thin partial absorber is mathematically indistinguishable from the change in the partial covering fraction \citep[e.g.,][]{miy12,mizu14}, 
the two partial covering fractions can  be the same by adjusting the column densities.
This model is called the variable double partial covering (VDPC) model \citep{mizu14}, and
has been successful to explain  spectral variations of a dozen of 
Seyfert galaxies  \citep{miy12,iso16,yam16,mido21}.
Hereafter the partial covering fraction is denoted as  $\alpha$, where $0 \le \alpha \le  1$.

The model we used is expressed as follows:   
\begin{equation} 
\begin{split}
F = &\lbrace (1 - \alpha + \alpha * {\tt W_1})(1 - \alpha + \alpha * {\tt W_2}) \\
&({\tt powerlaw} + {\tt const_{ref}} * {\tt relxill} + {\tt diskbb}) * {\tt W_3} * {\tt W_4} \\
&+ {\tt pexmon} + {\tt const_{Mn}} * {\tt G_{Mn}} \rbrace \\
&* {\tt const_{cross}}  *  {\tt zTBabs} * {\tt TBabs}.
\end{split}
\label{eq:VDPC}
\end{equation}

In the following, we explain the individual spectral components:

(1) {\tt powerlaw}:  The intrinsic X-ray component from the central X-ray source is modeled with a power-law function in the
concerned energy range (0.3 -- 79 keV).

(2)  ${\tt const_{ref}} * {\tt relxill}$: The   inner disk reflection is modeled using 
 {\tt relxill} v.2.1 \citep{gadau14,dauga14,dau22}.
The black hole spin parameter $a$ is fixed at 0 and the emissivity index is set to 3, which are the nominal values for the Schwarzschild black hole.  The inner and outer disk radius are $6\,R_g$ and $400\,R_g$, respectively, where $R_g$ is the gravitational radius.
The {\tt relxill} normalization 
is made equal to the {\tt powerlaw} one. 
The reflection rate relative to the direct X-ray emission 
(${\tt const_\text{ref}}$) and
the ionization parameter of the disk 
are  free parameters common to  all the spectra. 
The inclination angle with respect to the line-of-sight is assumed  to be 60 deg, since  Mrk 766 is classified as a ``complex'' NLS1 
which tends to  have a high inclination angle \citep{gar15}.
The cut-off energy, Fe abundance, and redshift of  {\tt relxill} are fixed at 1,000 keV,  the solar abundance, and the cosmic redshift, respectively (Table \ref{tab:fixpara}).

(3) {\tt pexmon}: 
A narrow emission line at 6.4 keV  corresponding to the fluorescent Fe-K line  was observed within all the spectra.
Therefore we introduced {\tt pexmon}, which represents the reflection by a distant neutral disk or torus \citep{mag95,nan07}.
The photon index and normalization are set equal to those of {\tt powerlaw} component, and the inclination, abundance, cut-off energy, and redshift are fixed at  those of the {\tt relxill} model. 
The only free parameter in this model is the reflection rate, which is tied across all the spectra.
The reflection rate corresponds to $\Omega / 2\pi$, where $\Omega$ is the
solid angle of the reflector.

(4) {\tt diskbb}: To explain the soft excess component, we adopted the  phenomenological disk blackbody model,  {\tt diskbb} \citep{mitu84,maki86}. 
The normalization is dimensionless value given by $\left({R_\mathrm{in}}/{D_{10}}\right)^2 \cos \theta$,
where $R_\mathrm{in}$ is the inner disk radius in km, $D_{10}$ is the distance to the X-ray source in 10 kpc, and $\theta$ is the inclination angle of the disk.
The normalization is allowed to vary independently for each spectrum.
The inner disk temperature ($T_\mathrm{in}$) is tied across all the spectra.

(5) {\tt TBabs} and {\tt zTBabs}:
The Galactic \ion{H}{I} column density toward Mrk 766  is $1.85\times 10^{20}$~cm$^{-2}$ (LAB map; \citealt{LAB}), which is modelled by  {\tt TBabs} \citep{ver96,wil00}.
The host galaxy absorption is modeled by {\tt zTBabs}, whose column density is a free parameter and the cosmic redshift is fixed.

(6) ${\tt W_i}$, where $i$ is from  1 to 4:
 Photo-ionized absorbers calculated with    MPI\_XSTAR \citep{dank18}, which  is an implementation of the photo-ionized simulation code {\tt xstar} (v.2.58e; \citealt{kal04})
incorporating the  Message Passing Interface (MPI).
The table models are calculated   assuming  a power-law incident spectrum ($\Gamma = 2.2$) and
discrete values of  the column density, ionization parameter, and turbulent velocity.
Table \ref{tab:xstar_cal} shows the parameter grids adopted.

${\tt W_1}$ and ${\tt W_2}$ are for  the double-layer partial covering clouds, and ${\tt W_3}$ represents 
the full-covering ionized warm absorber \citep{tom13}.
For these three absorbers, we use  the same table model, where 
the  line-of-sight velocity in the rest-frame and the turbulent velocity are fixed to null.
The ionization parameter was calculated between $10^{-2}$ and $10^5$ with 20 logarithmically spaced intervals, and the column density was calculated between 0.01 $\times 10^{22}$~cm$^{-2}$ and 1000 $\times 10^{22}$~cm$^{-2}$ with 20 logarithmically spaced intervals (400 grid points in total). 
Column densities and ionization parameters of each of the three absorbers are 
assumed to be invariable throughout the observational period.
In the double-layer absorber,
${\tt W_1}$ represents the higher-ionized layer while ${\tt W_2}$ is  for the lower-ionized core.
The covering fraction $\alpha$ is a free parameter for each spectrum.
The partial covering clouds are assumed to intervene  the central X-ray emitting components, namely, 
{\tt powerlaw} $+$ ${\tt const_{ref}}$ * {\tt relxill} $+$ {\tt diskbb}  in 
Eq. (\ref{eq:VDPC}), but not {\tt pexmon} that is considered to be located further from the black hole than the partial covering clouds.

${\tt W_4}$ represents the UFO, where  both the line-of-sight velocity and the turbulent velocity are  free parameters.
The ionization parameter was calculated between $10^3$ and $10^5$ with 20 logarithmically spaced intervals, and the column density was calculated between 0.01 $\times 10^{22}$~cm$^{-2}$ and 100 $\times 10^{22}$~cm$^{-2}$ with 20 logarithmically spaced intervals.  In addition,  the turbulent velocity was calculated between 100~km~s$^{-1}$ and 10,000~km~s$^{-1}$ with 5 logarithmically spaced intervals (2,000 grid points in total). 
All the ${\tt W_4}$ parameters are allowed to be free for each spectrum. 

(7) ${\tt const_{cross}}$ gives  a cross-calibration adjustment  between  XMM-pn and  \textit{NuSTAR} 
or between   \textit{Swift} and  \textit{NuSTAR}.

(8) ${\tt const_\text{Mn}} * G_\text{Mn}$: The Mn-K$\alpha$ line feature (\S\ref{sec:3.1}) is modeled by a positive Gaussian.
 Normalization of the line is assumed to be proportional to that 
 of {\tt powerlaw}.
 The ratio,   ${\tt const_\text{Mn}}$,
is tied for all the spectra.

\begin{table}
	\centering
	\caption{Fixed parameter in our spectral model.}
	\begin{tabular}{cccc} 
\\\hline
model   & parameter              & value\\\hline
{\tt TBabs}   & $N_{\rm H}$ ($10^{20}$cm$^{-2}$)              & 1.85  \\\hline
\multirow{7}{*}{\tt relxill} & Emissivity index          & 3                                \\
        & Spin parameter($a$)               & 0                                \\
        & Inclination~(degree)            & 60                           \\
        & $r_\text{in}$~($R_g$)             & 6                            \\
        & $r_\text{out}$~($R_g$)            & 400                            \\
        & Abundance        & Solar abundance$^{*1}$                            \\
        & Cut-off-Energy~(keV)            & 1000                        \\\hline
	\end{tabular}
 \label{tab:fixpara}
 \begin{threeparttable}
 \begin{tablenotes}
 \item[*1] \cite{gre98}
 \end{tablenotes}
\end{threeparttable}
\end{table}

\begin{table}
	\centering
	\caption{Parameter grids of the table models calculated  by  MPI\_XSTAR}
	\label{tab:xstar_cal}
\begin{threeparttable}
	\begin{tabular}{cccc}
\hline
parameter & ${\tt W_1}-{\tt W_3}$ & ${\tt W_4}$ \\\hline
$N_{\rm H}$ ($10^{22}$cm$^{-2}$)  & 0.01 - 1000 & 0.01 - 100 \\
$\log \xi$            & $-2$ - 5  & 3 - 5  \\
vturb (km s$^{-1}$)$^{*1}$ & 0 (fixed) & 100 - 10,000 \\\hline
number of grids$^{*2}$ & 400       & 2,000       \\\hline
\end{tabular}
\begin{tablenotes}
\item[*1] Turbulent velocity in the absorbing plasma.
\item[*2] $N_{\rm H}$, $\xi$ and vturb are logarithmically equally delimited by  20, 20, and 5,  respectively
\end{tablenotes}
\end{threeparttable}
\end{table}

\subsection{Spectral fitting} \label{sec3.3}

We conducted simultaneous spectral fitting using  the model in Eq.\ (\ref{eq:VDPC})
for all the 10 spectral datasets; XMM1--8, XMM9 $\&$ Nu11, and Nu10 $\&$ Sw12.
We try to find a solution that the disk geometry and properties of the partial and warm absorbers
exhibit minimal variations over the entire observation for  $\sim$15 years. Accordingly, we have fixed several parameters at
constant values (Table \ref{tab:fixpara}), tied some other parameters for all the ten datasets (Table \ref{tab:VDPC_result_para_common}), and allowed the remainder to vary for each dataset (Table \ref{tab:VDPC_result_para_free}).

\begin{figure*}
\centering
	\includegraphics[width=8cm]{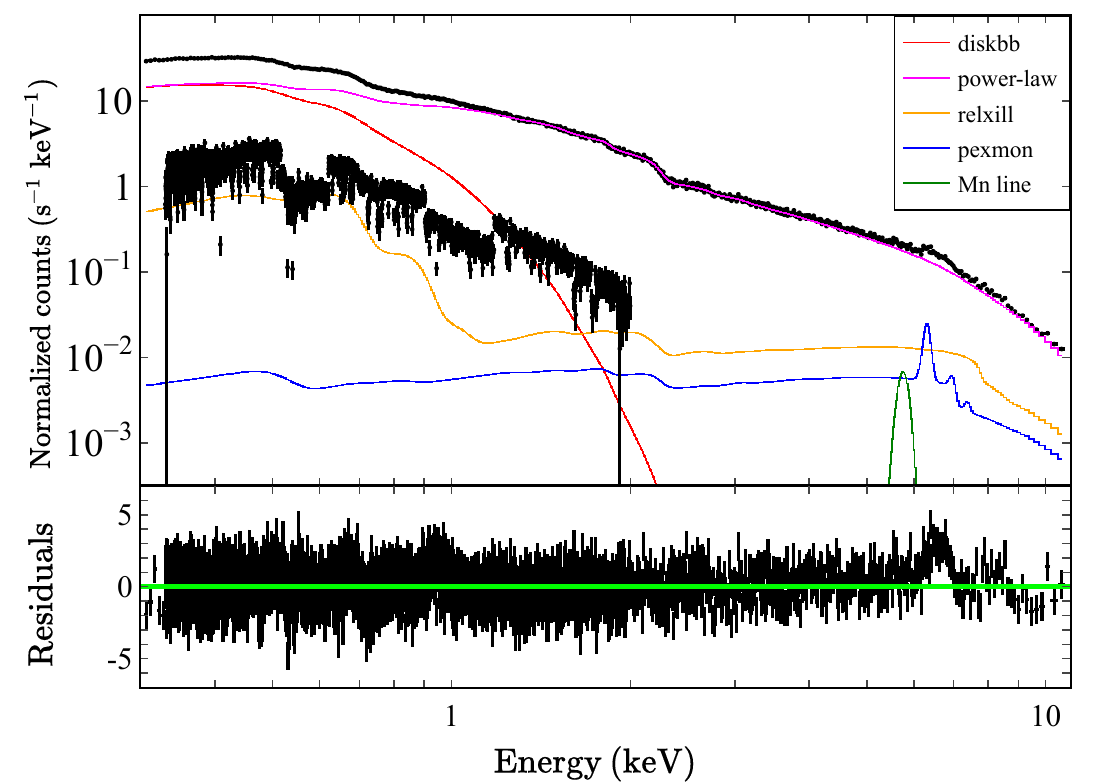}
	\includegraphics[width=8cm]{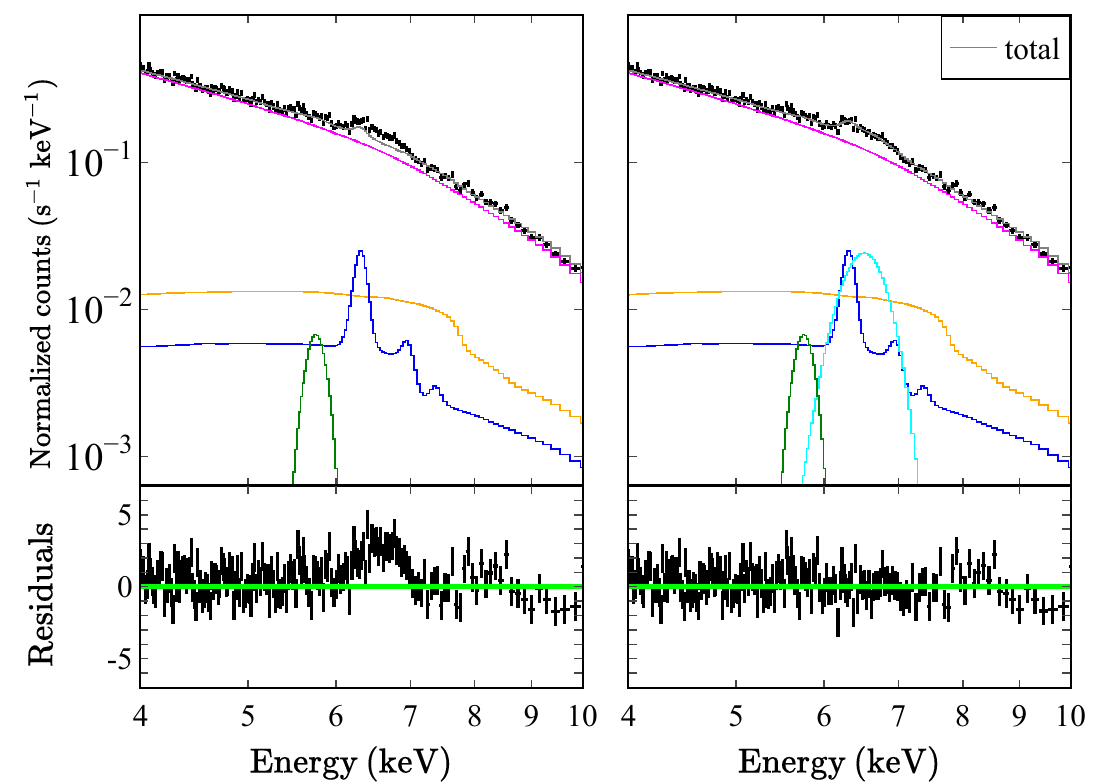}
	\includegraphics[width=8cm]{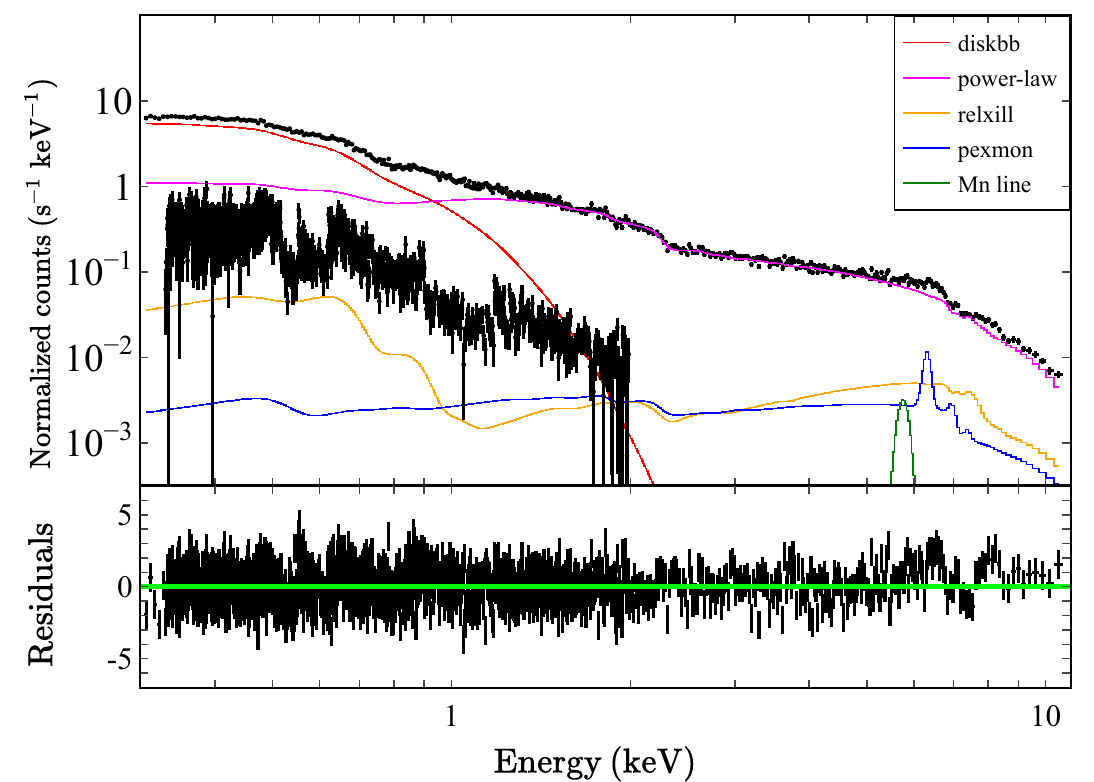}
	\includegraphics[width=8cm]{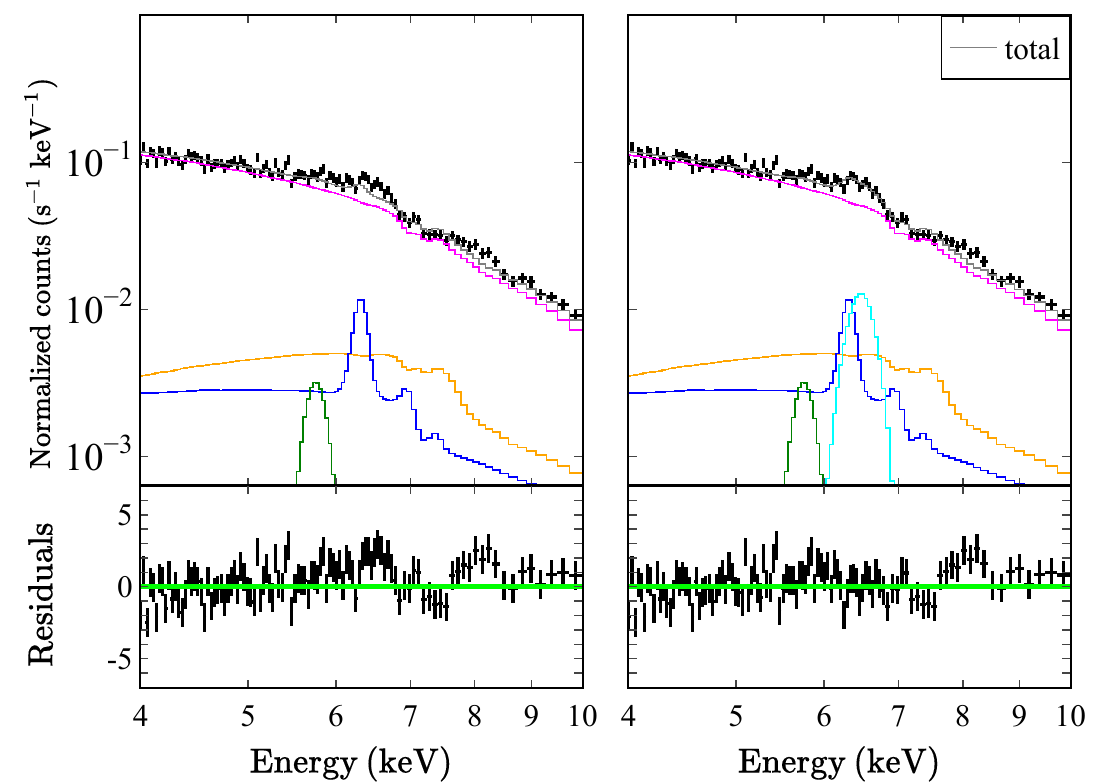}
	\caption{(left panels)  Spectra fitting results of XMM2 (upper) and XMM3 (lower) in 0.3--11~keV  with the model shown in Eq.\ (\ref{eq:VDPC}).
 The black shows the observed 
 spectra and residuals for  EPIC-pn and RGS, and the red,  
 magenta, orange, blue, green, and gray lines show the best-fit  EPIC-pn components  of {\tt diskbb},
 {\tt powerlaw}, {\tt relxill}, {\tt pexmon}, ${\tt G_\text{Mn}}$, and the total model spectrum, respectively. 
(center panels) Expansion of the left-panels in 4 -- 10 keV. (right panels) Fitting with an additional iron emission line ${\tt G_\text{Fe}}$ (cyan).}
    \label{fig:nogauss}
\end{figure*}

\begin{figure*}
\centering
\includegraphics[width=15cm, clip]{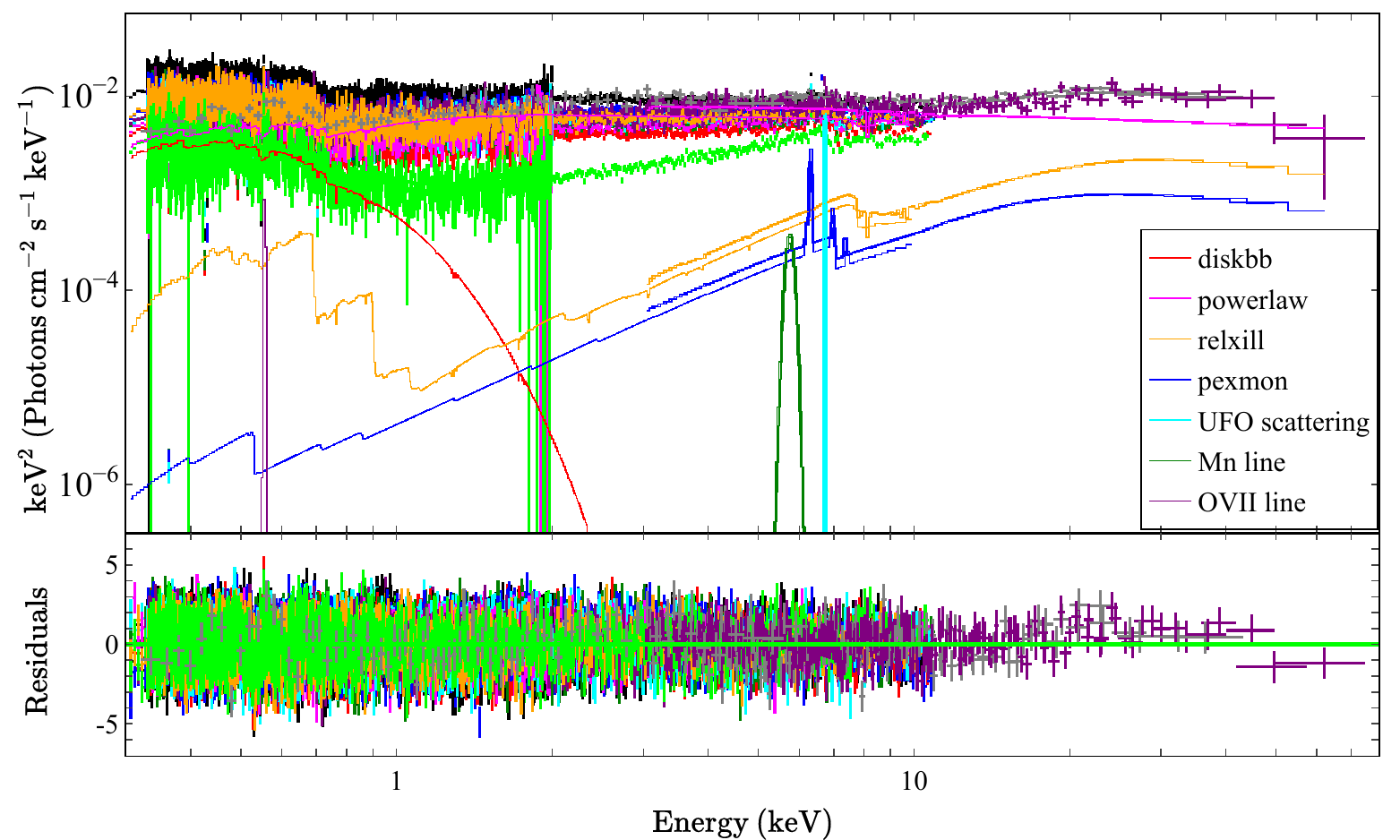}
   \caption{Result of the simultaneous spectra fitting of the 10 datasets with the double partial covering model in 0.3 -- 79 keV. 
   Data points and residuals of the 10 datasets,  XMM1--8, XMM9\&Nu11 and Nu10\&Sw12 are represented with different colors as follows;
   orange (XMM1),  black (XMM2),  green (XMM3), red (XMM4),  green (XMM5),  blue (XMM6),  cyan (XMM7),  magenta (XMM8),
   purple (XMM9 \& Nu11), and  gray  (Nu10 \& Sw12). Individual spectral model components (from diskbb to \ion{O}{VII} line) to fit  XMM9 \& Nu11
   are shown in different colors.
   }
  \label{fig:allfit_vdpc}
\end{figure*}

Fig. \ref{fig:nogauss} (left) shows fitting result of  XMM2 (the brightest one) and XMM3 (the dimmest one) in 0.3 -- 11 keV with the double partial covering model in Eq.\ (\ref{eq:VDPC}).
We found significant residuals at $\sim$$0.55$~keV and  6.4--6.7~keV.
The former is likely to be the \ion{O}{VII} line  from the photoionized plasma \citep{bui18},
for which we introduce a Gaussian line (${\tt G_\text{OVII}}$) for further analysis with 
its normalization and energy being tied for the 10 datasets.
Fig. \ref{fig:nogauss} (center) shows  expansion of the left-panels in 4 -- 10 keV
to demonstrate the   residual at 6.4--6.7~keV, which is more prominent in the brighter spectrum.
To explain this feature,  we add a  Gaussian line  (${\tt G_\text{Fe}}$)  (Fig. \ref{fig:nogauss} right), whose origin will be discussed in \S\ref{sec:Gorigin}. 
The normalization, energy, and intrinsic width of ${\tt G_\text{Fe}}$ are made free parameters for
each dataset.
We also remark that the UFO absorption feature and the partial covering feature
(manifested by the spectral flattening below $\sim$6 keV)
are more prominent in the dimmer spectrum, as 
suggested by \citet{Tur07} and \citet{ris11}.
The simultaneous fitting results of the 10 datasets in 0.3 -- 79 keV are shown in Fig.\ \ref{fig:allfit_vdpc} and Tables \ref{tab:VDPC_result_para_common} and \ref{tab:VDPC_result_para_free}.

%%%%%%%%%%%%%%%%%%%%%%%%%%%%%VDPC#######################

\begin{table}
    \centering
    \caption{Fitting results of the tied parameters in the double-layer partial covering model}
    \label{tab:VDPC_result_para_common}
    \begin{threeparttable}
    \begin{tabular}{|c|c|c|}
    \hline
        model & paramter &value   \\ \hline
        {\tt zTBabs} & $N_{\rm{H}} (10^{20} \rm{cm}^{-2})$ &$ 1.53 ^{+ 0.02 }_{ -0.02 }$   \\ \hline
        \multirow{2}{*}{W$_1$} & $N_{\rm{H}} (10^{23} \rm{cm}^{-2})$ &  $ 3.89 ^{+ 0.08 }_{ -0.08 } $ \\
        & $\log \xi$ &  $ 1.68 ^{+ 0.01 }_{ -0.02 } $ \\ \hline
         \multirow{2}{*}{W$_2$} & $N_{\rm{H}} (10^{22} \rm{cm}^{-2})$ & $ 2.10 ^{+ 0.03 }_{ -0.03 } $  \\
        & $\log \xi$ &  $ < -2$ \\ \hline
        \multirow{2}{*}{W$_3$} & $N_{\rm{H}} (10^{21} \rm{cm}^{-2})$ & $ 2.87^{+0.01 }_{ -0.01 } $\\
        & $\log \xi$ &$ 0.478 ^{+ 0.004 }_{ -0.004 } $   \\\hline
        {\tt diskbb} & $T_\mathrm{in}$ (keV)&  $ 0.1316 ^{+ 0.0001 }_{ -0.0001 } $ \\ \hline
        {\tt powerlaw} & $\Gamma$ &  $ 2.24 ^{+ 0.02 }_{ -0.01 } $ \\ \hline
        {\tt relxill} & $\log \xi$ & $ 0.30 ^{+ 0.01 }_{ -0.03 } $
  \\ \hline
        ${\tt const_\text{ref}}$ & {\tt powerlaw} ratio & $ 0.0090 ^{+ 0.0003 }_{ -0.0004 } $\\ \hline
        {\tt pexmon} & Reflection rate & $ -0.26 ^{+ 0.02 }_{ -0.02 } $  \\ \hline 
        \multirow{3}{*} {${\tt G_\text{OVII}}$} & Energy (keV) &$ 0.5543 ^{+ 0.0000 }_{ -0.0002 } $  \\
         & $\sigma (\rm{keV})$ & $ 0.001 ^{+ 0.001 }_{ -0.000 } $ \\
         & normalization ($10^{-5}$ photons s$^{-1}$ cm$^{-2}$) & $ 2.4 ^{+ 0.3 }_{ -0.7 } $ \\\hline
         \multirow{2}{*} {${\tt G_\text{Mn}}$} & Energy (keV) & $ 5.76 ^{+ 0.05 }_{ -0.07 } $  \\
         & $\sigma (\rm{keV})$ & $ 0.09 ^{*1}$ \\\hline
         ${\tt const_\text{Mn}}$ & {\tt powerlaw} ratio & $ 0.0002 ^{+ 0.0001 }_{ -0.0001 } $  \\ \hline        
    \end{tabular}
\begin{tablenotes}
\item[*1] not constrained
\end{tablenotes}
\end{threeparttable}
\end{table}

\begin{table*}
    \centering
    \caption{Fitting results of the free parameters with the double-layer partial covering model}
    \label{tab:VDPC_result_para_free}
    \begin{threeparttable}
    \begin{tabular}{|c|c|c|c|c|c|c|c|}
    \hline
        model & parameter & XMM1 & XMM2 & XMM3 & XMM4 & XMM5  \\\hline
        W$_1$\&W$_2$ & $\alpha$ & $ 0.186 ^{+ 0.005 }_{ -0.001 } $& $ < 0.000$ & $ 0.643 ^{+ 0.001 }_{ -0.001} $ & $ 0.270 ^{+ 0.001 }_{ -0.001 } $ & $ 0.195 ^{+ 0.001 }_{ -0.001 } $
  \\\hline
        \multirow{4}{*}{W$_4$} & $N_{\rm{H}} (10^{23} \rm{cm}^{-2})$ &
         $ 0.8 ^{+ 0.2 }_{ -0.2 } $  
        & $ < 1.8 $ & $ 0.89 ^{+ 0.34 }_{ -0.30 } $ &$ 0.61 ^{+ 0.16 }_{ -0.15} $ &  $ 2.06 ^{+ 0.97 }_{ -0.69 } $    \\
         & $\log \xi$ & $ 3.73 ^{+ 0.06 }_{ -0.04 } $ & $ > 4.2$ & $ 3.89 ^{+ 0.15 }_{ -0.11 }$ & $ 3.86 ^{+ 0.09}_{ -0.06 } $ & $ 4.11 ^{+ 0.16 }_{ -0.10 } $\\
         & vturb ($10^{3}$km s$^{-1}$) & 
          $ 1.4 ^{+ 1.4 }_{ -0.7 } $ 
         & $ 0.3 ^{*1}$ & $ 0.6 ^{+ 2.1 }_{ -0.5 }$ & $ > 7.0 $& $ < 0.1 $    \\
         & z & $ -0.100 ^{+ 0.002 }_{ -0.002} $ & $ -0.271$(fixed) & $ -0.038 ^{+ 0.003 }_{ -0.008 }$ & $ -0.065 ^{+ 0.003 }_{ -0.003 }$ & $ -0.066 ^{+ 0.003 }_{ -0.001 }$ \\\hline
        {\tt diskbb} & normalization ($10^{3}$) & $ 7.81 ^{+ 0.06 }_{ -0.06 } $
 & $ 7.89 ^{+ 0.03 }_{ -0.03 }$ & $ 13.6 ^{+ 1.3 }_{ -1.0 } $ & $ 5.90 ^{+ 0.03 }_{ -0.03 }$ & $ 5.88 ^{+ 0.03 }_{ -0.03 } $    \\\hline 
        {\tt powerlaw} & normalization ($10^{-2}$ photons s$^{-1}$ cm$^{-2}$ keV$^{-1}$) & $ 0.907 ^{+ 0.003 }_{ -0.003 } $ & $ 1.156 ^{+ 0.002 }_{ -0.002 } $ & $ 0.560 ^{+ 0.003 }_{ -0.003 }$ & $ 0.705 ^{+ 0.002 }_{ -0.002 } $ & $ 0.815 ^{+ 0.002 }_{ -0.002 } $   \\\hline
        \multirow{3}{*}{${\tt G_\text{Fe}}$} & Energy (keV) & $ 6.61 ^{+ 0.14 }_{ -0.15 } $& $ 6.53 ^{+ 0.05 }_{ -0.05} $ & $ 6.48^{+ 0.06 }_{ -0.07 } $ & $ 6.40 ^{+ 0.07 }_{ -0.07 }$ & $ 6.68 ^{+ 0.10 }_{ -0.10 } $\\
         & $\sigma (\rm{keV})$ & $ 0.17 ^{+ 0.15 }_{ -0.09 } $ & $ 0.28 ^{+ 0.05}_{ -0.04 } $& $ 0.16 ^{+ 0.12 }_{ -0.05 } $ & $ 0.13 ^{+ 0.07 }_{ -0.05 } $ & $ 0.17 ^{+ 0.10 }_{ -0.07 } $\\
         & normalization ($10^{-5}$ photons s$^{-1}$ cm$^{-2}$) & $ 0.79 ^{+ 0.43 }_{ -0.42 } $&  $ 2.61 ^{+ 0.35 }_{ -0.34 } $ & $ 0.84 ^{+ 0.21 }_{ -0.20 }$ & $ 0.68 ^{+ 0.21 }_{ -0.21 }$ & $ 0.69 ^{+ 0.25 }_{ -0.25 } $\\\hline
         \hline
         \multicolumn{2}{c}{$\chi ^2$/d.o.f} &1.13&1.32& 1.34& 1.13&1.18\\\hline
    \hline     
        model & parameter & XMM6 & XMM7 & XMM8 & XMM9\&Nu11 &   Nu10\&Sw12     \\\hline
        W$_1$\&W$_2$ & $\alpha$ &  $ 0.186 ^{+ 0.004 }_{ -0.001 }$ & $ 0.180 ^{+ 0.001 }_{ -0.001 }$ & $ 0.226 ^{+ 0.002 }_{ -0.002 } $ & $ 0.128 ^{+ 0.018 }_{ -0.019 } $ & $ 0.088 ^{+ 0.010}_{ -0.010 } $   \\\hline
        \multirow{4}{*}{W$_4$} & $N_{\rm{H}} (10^{23} \rm{cm}^{-2})$ & $ 1.05 ^{+ 0.37 }_{ -0.26 } $&$ 0.93 ^{+ 0.32 }_{ -0.29 } $ & $ 0.51 ^{+ 0.19 }_{ -0.19 } $ &  $ > 5.3$ &$ 7.6 ^{+ 2.1 }_{ -1.8 } $   \\
         & $\log \xi$ & $ 4.27 ^{+ 0.13 }_{ -0.12 } $&$ 4.08 ^{+ 0.13 }_{ -0.09 }$ & $ 3.77 ^{+ 0.13 }_{ -0.08 } $ & $ 4.66 ^{+ 0.26 }_{ -0.19 } $ & $ 4.46 ^{+ 0.14 }_{ -0.56 } $  \\
         & vturb ($10^{3}$km s$^{-1}$) & $ 3.2 ^{+ 4.3 }_{ -1.4 } $&$ 3.8 ^{+ 5.0 }_{ -2.5 } $ & $ > 0.69$ & $ < 0.8 $ & $ < 0.28 $  \\
         & z &  $ -0.074 ^{+ 0.005 }_{ -0.008 } $ &$ -0.052 ^{+ 0.002 }_{ -0.002 }$ & $ -0.072 ^{+ 0.009 }_{ -0.011 } $
 & $ -0.145 ^{+ 0.002 }_{ -0.002 }$ & $ -0.080 ^{+ 0.010 }_{ -0.014 }$    \\\hline
        {\tt diskbb} & normalization ($10^{3}$) &$ 6.97 ^{+ 0.04 }_{ -0.04 } $& $ 5.58 ^{+ 0.09 }_{ -0.04 } $ & $ 5.2 ^{+ 0.06 }_{ -0.06 }$ & $ 4.50 ^{+ 0.09 }_{ -0.08 }$ &  $ < 0.24 $   \\\hline
        {\tt powerlaw} & normalization ($10^{-2}$ photons s$^{-1}$ cm$^{-2}$ keV$^{-1}$) & $ 1.015 ^{+ 0.002 }_{ -0.002 } $& $ 0.857 ^{+ 0.002 }_{ -0.002 }$ & $ 0.794 ^{+ 0.004 }_{ -0.004 } $& $ 0.914 ^{+ 0.005 }_{ -0.005 } $ & $ 1.094 ^{+ 0.031 }_{ -0.007 } $
   \\\hline
        \multirow{3}{*}{${\tt G_\text{Fe}}$} & Energy (keV) & $ 6.64 ^{+ 0.03 }_{ -0.04 } $& $ 6.54 ^{+ 0.05 }_{ -0.05 } $ & $ > 6.83 $ & $ 6.75 ^{+ 0.09 }_{ -0.11 }$ &{-}    \\
         & $\sigma(\rm{keV})$ &$ < 0.11$  & $ < 0.17$ & $ 0.03 ^{*1}$ & $ 0.001 ^{*1}$ & {-}    \\
         & normalization ($10^{-5}$ photons s$^{-1}$ cm$^{-2}$) & $ 0.47 ^{+ 0.19 }_{ -0.19 } $& $ 0.42 ^{+ 0.18 }_{ -0.18 } $ & $ 0.37 ^{+ 0.32}_{ -0.32 }$ & $ 0.80 ^{+ 0.41 }_{ -0.41 }$ & {-}    \\\hline
 \multirow{4}{*}{${\tt const_\text{cross}}$} & FPMA/XMM-pn & -& - & - &  $ 1.28 ^{+ 0.02 }_{ -0.02 } $ & {-}   \\
        ~ & FPMB/XMM-pn & -& - & - &  $ 1.24 ^{+ 0.02 }_{ -0.02 } $ & {-}   \\
        ~ & FPMA/Swift-XRT & -& - & - & {-}  &  {$ 1.25 ^{+ 0.03 }_{ -0.04 }$} \\
        ~ & FPMB/Swift-XRT & -& - & - & {-}  &   $ 1.21 ^{+ 0.01}_{ -0.01 } $
 \\\hline
        \hline
 \multicolumn{2}{c}{$\chi ^2$/d.o.f} & 1.14& 1.13&1.01& {1.01} &{1.54}\\\hline  
    \end{tabular}
    \begin{tablenotes}
\item[*1] not constrained
\end{tablenotes}
    \end{threeparttable}
\end{table*}

\subsection{The triple partial covering model}

\begin{figure*}
\centering
	\includegraphics[width=14cm]{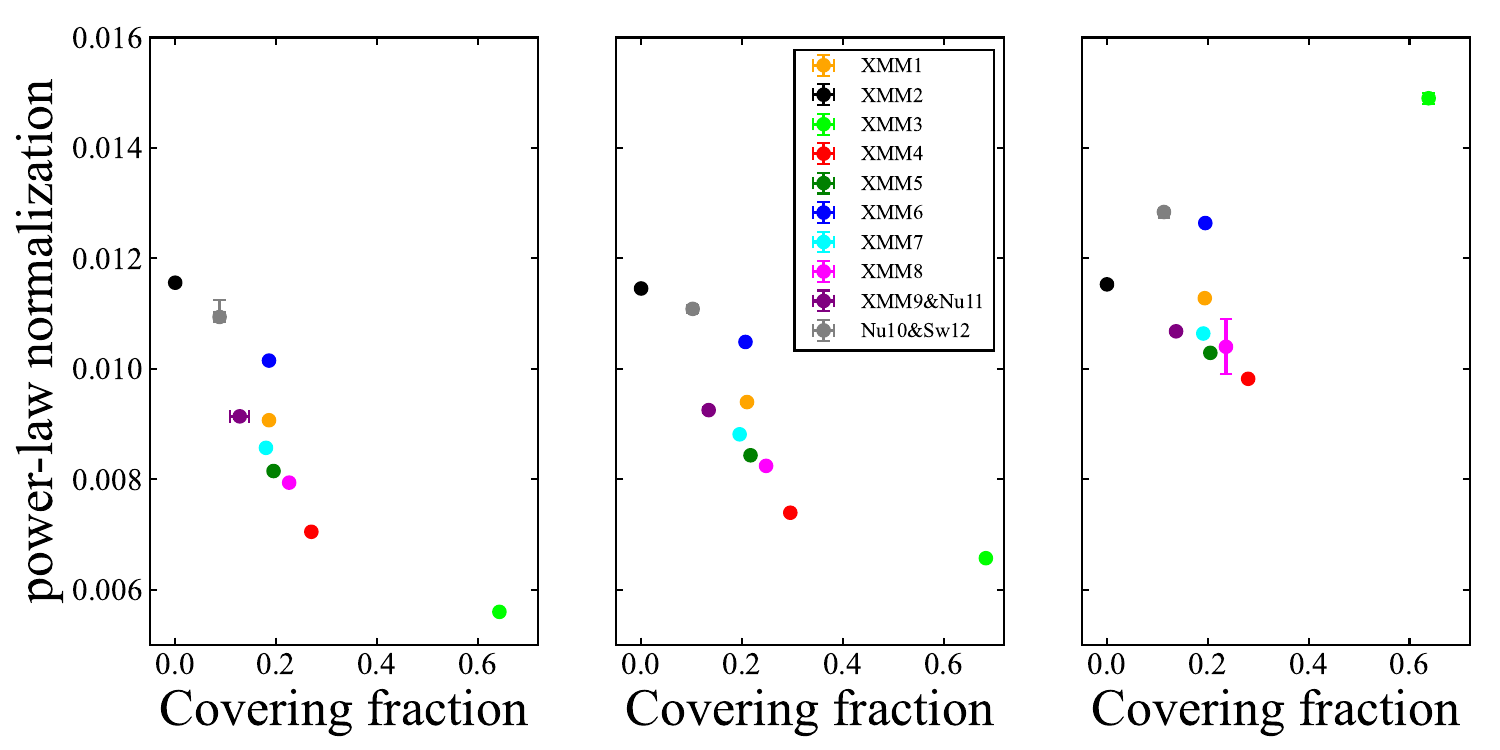}
    \caption{Relation between the partial covering fraction and the power-law normalization. The left panel is the result of 
    the double partial covering  model, the center panel is the result of adding the {\tt cabs} model, and the right panel is the result of the  triple partial covering model. The 
    datasets are distinguished by different colors.}
    \label{fig:Cf-norm}
\end{figure*}

As shown in Fig.\ \ref{fig:allfit_vdpc} and Table \ref{tab:VDPC_result_para_free}, the double partial covering model
can explain all the 10 spectral datasets taken over $\sim$15 years with most significantly variable parameters
being the power-law normalization and the partial covering fraction.
The left panel of Fig. \ref{fig:Cf-norm} displays  the power-law normalization versus the covering fraction
for the 10 datasets, which indicates a negative correlation between the two parameters. This negative correlation is unexpected, 
since the intrinsic X-ray luminosity, which is supposed to be proportional to the power-law normalization, and the geometrical covering fraction of the
X-ray emitting region by the outer clumps should not be  directly connected.

First, we examine whether the electron scattering  of  X-rays 
out of the line of sight due to the partial covering clouds
 could explain this correlation.
We add 
the {\tt cabs} model for the electron scattering to the partial absorber, where the column density is fixed to the sum of the two partial covering clouds. 
Even in this case, the negative correlation does not disappear  (the center panel of Fig. \ref{fig:Cf-norm}).

Next, we propose the presence  of the third partial covering layer which is
completely opaque \citep[e.g.,][]{miy12}.  In this case,  
as the partial covering fraction increases,  the intrinsic X-rays are more blocked, and the power-law normalization  decreases apparently,
even if the intrinsic X-ray luminosity does not vary.
Thus, we adopt the ''triple partial covering model", which extends the double-layer partial covering model 
by including an additional neutral core (${\tt W_5}$). The third partial covering layer is modeled using 
{\tt zpcfabs}  with  the abundance by \citet{wil00}.
We can express the triple partial covering model as
\begin{equation}
\begin{split}
F = &\lbrace (1 - \alpha + \alpha * {\tt W_5})(1 - \alpha + \alpha * {\tt W_1})(1 - \alpha + \alpha * {\tt W_2}) \\
&({\tt powerlaw} + {\tt const_{ref}} * {\tt relxill} + {\tt diskbb}) * {\tt W_3} * {\tt W_4} \\
&+ {\tt pexmon} + {\tt const_{Mn}} * {\tt G_{Mn}} + {\tt G_{Fe}} + {\tt G_{OVII}} \rbrace \\
& * {\tt const_{cross}}  *  {\tt zTBabs} * {\tt TBabs}.
\end{split}
\end{equation}

\begin{figure*}
\centering
	\includegraphics[width=15cm, clip]{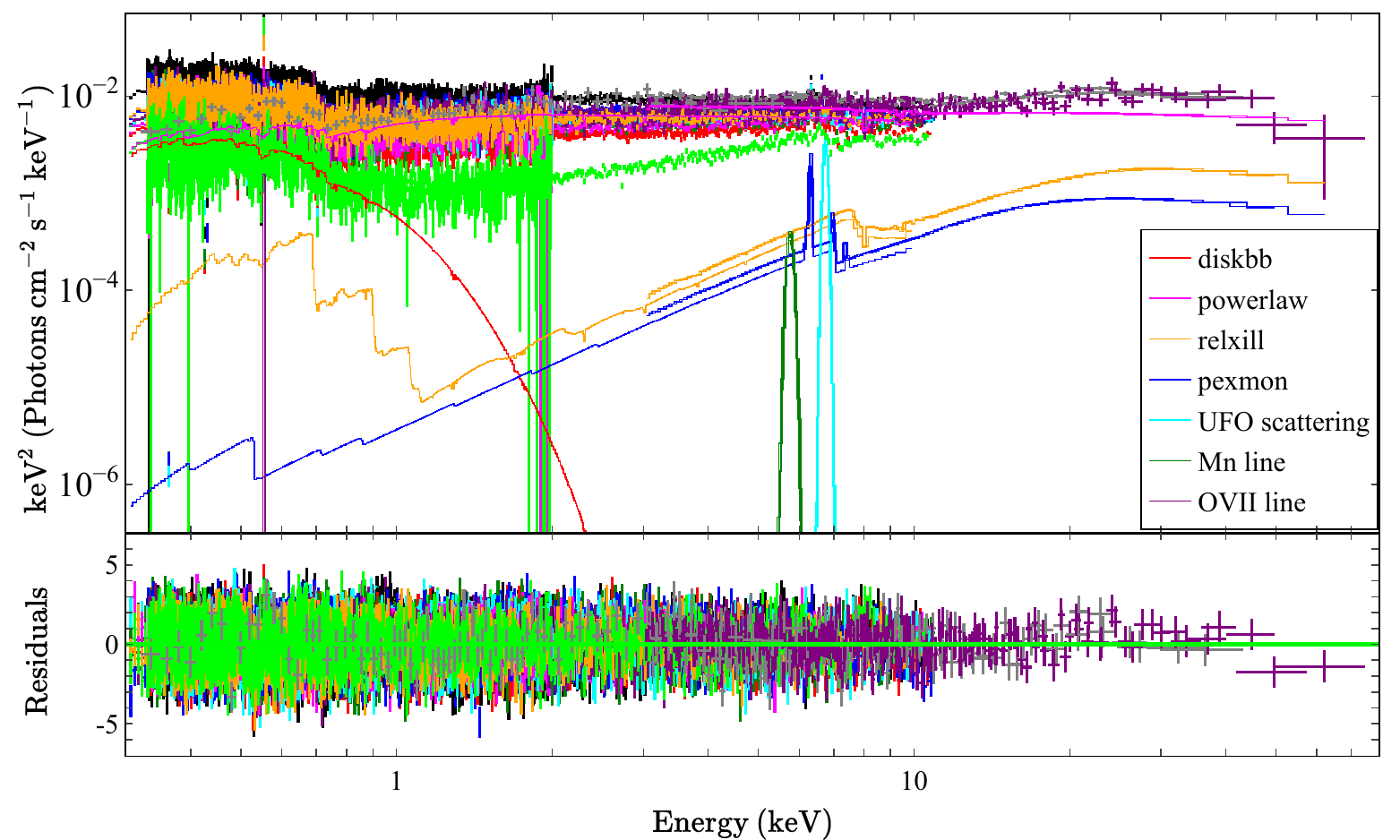}
    \caption{Result of the simultaneous spectra fitting of the 10 datasets with  the triple partial covering model in 0.3-79 keV. 
    See the caption of Fig.\ \ref{fig:allfit_vdpc} for other explanation.}
    \label{fig:allfit}
\end{figure*}
We can explain all the data with the triple partial covering model 
(Fig.\ \ref{fig:allfit}, Tables \ref{tab:triple_result_para_common} and \ref{tab:triple_result_para_free}) as well  as with the double partial covering  model. 
The right panel of Fig.\ \ref{fig:Cf-norm} shows the relation between the power-law normalization and the covering fraction
in the triple partial covering model, which does not show  the obvious correlation.
Therefore, we consider that the triple partial covering model is more preferable to the double-layer partial covering model.

%%%%%%%%%%%%%%%%%%%%%%%%%%%%%triple#######################

\begin{table}
    \centering
    \caption{Fitting results of the tied parameters with the triple-layer partial covering model.}
    \label{tab:triple_result_para_common}
    \begin{threeparttable}
    \begin{tabular}{|c|c|c|}
    \hline
        model & paramter &value   \\ \hline
        {\tt zTBabs} & $N_{\rm{H}} (10^{20} \rm{cm}^{-2})$ & $ 1.67 ^{+ 0.02 }_{ -0.00 } $\\ \hline        
        \multirow{2}{*}{W$_1$} & $N_{\rm{H}} (10^{23} \rm{cm}^{-2})$ &  $ 3.89 ^{+ 0.08 }_{ -0.08 } $ \\
        & $\log \xi$ &  $ 1.70 ^{+ 0.01 }_{ -0.01 } $\\ \hline
        \multirow{2}{*}{W$_2$} & $N_{\rm{H}} (10^{22} \rm{cm}^{-2})$ & $ 1.79 ^{+ 0.03 }_{ -0.02 } $  \\
        & $\log \xi$ &  $ < -2 $ \\ \hline
        \multirow{2}{*}{W$_3$} & $N_{\rm{H}} (10^{21} \rm{cm}^{-2})$ & $ 2.79 ^{+0.01 }_{ -0.02 } $  \\
        & $\log \xi$ &$ 0.474 ^{+ 0.004 }_{ -0.004 } $   \\\hline
        W$_5$ & $N_{\rm{H}} (10^{24} \rm{cm}^{-2})$ & $ 4.9 ^{+ 1.5 }_{ -0.9 } $  \\ \hline 
        {\tt diskbb} & $T_\mathrm{in}$ (keV) &  $ 0.1291 ^{+ 0.0001 }_{ -0.0001} $ \\ \hline
        {\tt powerlaw} & $\Gamma$ &  $ 2.210 ^{+ 0.001 }_{ -0.001 } $ \\ \hline
        {\tt relxill} & $\log \xi$ & $ 0.49 ^{+ 0.02 }_{ -0.04 } $  \\ \hline
        ${\tt const_\text{ref}}$ & {\tt powerlaw} ratio & $ 0.0063 ^{+ 0.0002 }_{ -0.0003 } $  \\ \hline
        {\tt pexmon} & Reflection rate & $ -0.20 ^{+ 0.01 }_{ -0.01 } $  \\ \hline
        \multirow{3}{*} {${\tt G_\text{OVII}}$} & Energy (keV) &$ 0.5543 ^{+ 0.0001 }_{ -0.0004 } $  \\
         & $\sigma (\rm{keV})$ & $ < 0.0004 $ \\
         & normalization ($10^{-5}$ photons s$^{-1}$ cm$^{-2}$) & $ 2.8 ^{+ 0.6 }_{ -0.3 } $ \\\hline
         %Ebi: lineのnormalizationの単位を書いてください。photons/s/cm^2でしょうか？
         \multirow{2}{*} {${\tt G_\text{Mn}}$} & Energy (keV) & $ 5.76 ^{+ 0.05 }_{ -0.07 } $  \\
         & $\sigma (\rm{keV})$ & $0.07^{*1}$ \\\hline
         ${\tt const_\text{Mn}}$ & {\tt powerlaw} ratio & $ 0.0002 ^{+ 0.0001 }_{ -0.0001 } $  \\ \hline
    \end{tabular}
\begin{tablenotes}
\item[*1] not constrain
\end{tablenotes}
\end{threeparttable}
\end{table}

\begin{table*}
    \centering
    \caption{Fitting results of the free parameters with the triple-layer partial covering model}
    \label{tab:triple_result_para_free}
    \begin{threeparttable}
    \begin{tabular}{|c|c|c|c|c|c|c|c|}
    \hline
        model & parameter & XMM1 & XMM2 & XMM3 & XMM4 & XMM5   \\\hline
        W$_1$\&W$_2$\&W$_5$ & $\alpha$ & $ 0.194 ^{+ 0.001 }_{ -0.001 } $ & $< 0.000$ & $ 0.638 ^{+ 0.001 }_{ -0.000 } $ & $ 0.280 ^{+ 0.001 }_{ -0.001 } $ & $ 0.205 ^{+ 0.001 }_{ -0.001 } $ \\\hline
        \multirow{4}{*}{W$_4$} & $N_{\rm{H}} (10^{23} \rm{cm}^{-2})$ &$ 0.70 ^{+ 0.18}_{ -0.13 } $ & $< 0.99$ & $ 0.83 ^{+ 0.17 }_{ -0.19 }$ & $ 1.08 ^{+ 0.49 }_{ -0.48 } $ & $ 2.8 ^{+ 1.5 }_{ -1.2 } $ \\
        ~ & $\log \xi$ & $ 3.71 ^{+ 0.06 }_{ -0.05} $ & $ 5.0^{*1}$ & $ 3.68 ^{+ 0.18 }_{ -0.15 }$ & $ 3.96 ^{+ 0.12 }_{ -0.10 } $ & $ 4.29 ^{+ 0.21 }_{ -0.14 }$   \\
        ~ & vturb ($10^{3}$~km s$^{-1}$) & $ 1.2 ^{+ 1.6 }_{ -0.5 } $ & $ 0.1^{*1}$ & $ 5.3^{+ 3.5 }_{ -3.5 }$ & $ < 0.62 $ & $ < 0.19$ \\
        ~ & z & $ -0.101 ^{+ 0.002 }_{ -0.002 } $ & $ -0.271$(fixed) & $ -0.038 ^{+ 0.003 }_{ -0.003 }$ & $ -0.065 ^{+ 0.001 }_{ -0.002 }$ & $ -0.066 ^{+ 0.003 }_{ -0.002 } $   \\\hline
        {\tt diskbb} & normalization ($10^{3}$) & $ 11.28 ^{+ 0.07 }_{ -0.09 } $ & $ 8.87 ^{+ 0.03 }_{ -0.03 } $& $ 55.0 ^{+ 0.3 }_{ -0.3 }$ & $ 9.57 ^{+ 0.05 }_{ -0.06 } $ & $ 8.64 ^{+ 0.04 }_{ -0.05 } $  \\\hline
        {\tt powerlaw} & normalization ($10^{-2}$ photons s$^{-1}$ cm$^{-2}$ keV$^{-1}$) & $ 1.128 ^{+ 0.004 }_{ -0.004 } $& $ 1.153 ^{+ 0.003 }_{ -0.001}$& $ 1.49 ^{+ 0.01 }_{ -0.01 }$ & $  0.982 ^{+ 0.003 }_{ -0.003 } $ & $ 1.029 ^{+ 0.002 }_{ -0.003 } $   \\\hline
        \multirow{3}{*}{${\tt G_\text{Fe}}$} & Energy (keV)& $ 6.60 ^{+ 0.14 }_{ -0.14} $ & $ < 6.53 $ & $ 6.56 ^{+ 0.05 }_{ -0.05 } $ & $ 6.41 ^{+ 0.07 }_{ -0.07 } $ & $ 6.66 ^{+ 0.09 }_{ -0.10 } $  \\
        ~ & $\sigma~(\rm{keV})$ & $ 0.19 ^{+ 0.15 }_{ -0.09 } $ & $ 0.42 ^{+ 0.08 }_{ -0.06 }$ & $ < 0.13$ & $ 0.13 ^{+ 0.07 }_{ -0.05 } $ & $ 0.19 ^{+ 0.09 }_{ -0.07 } $  \\
        ~ & normalization ($10^{-5}$ photons s$^{-1}$ cm$^{-2}$) & $ 0.98 ^{+ 0.45}_{ -0.45 } $ & $ 3.97 ^{+ 0.41 }_{ -0.41 } $& $ 0.52 ^{+ 0.17 }_{ -0.16 }$ & $ 0.71 ^{+ 0.22}_{ -0.22}$ & $ 0.81 ^{+ 0.26 }_{ -0.26 } $  \\\hline
        \hline
        \multicolumn{2}{c}{$\chi ^2$/d.o.f} &1.13&1.32& 1.34& 1.13&1.17\\\hline
    \hline
        model & parameter & XMM6 & XMM7 & XMM8 & XMM9\&Nu11 & Nu10\&Sw12  \\\hline
        W$_1$\&W$_2$\&W$_5$ & $\alpha$ & $ 0.195 ^{+ 0.001 }_{ -0.001 } $&$ 0.191 ^{+ 0.001 }_{ -0.001 }$&  $ 0.237 ^{+ 0.001 }_{ -0.001 } $
 & $ 0.137 ^{+ 0.002 }_{ -0.002 } $& {$ 0.113 ^{+ 0.004 }_{ -0.003 }$}  \\\hline
        \multirow{4}{*}{W$_4$} & $N_{\rm{H}} (10^{23} \rm{cm}^{-2})$ & $ 1.63 ^{+ 0.87 }_{ -0.61 } $ & $ 1.02 ^{+ 0.60 }_{ -0.38 }$ & $ 0.36 ^{+ 0.55 }_{ -0.14 } $ & $> 5.3 $& $ > 6.9 $  \\
        ~ & $\log \xi$ & $ 4.45 ^{+ 0.20 }_{ -0.15 } $& $ 4.13^{+ 0.17 }_{ -0.13 }$ & $ 3.77 ^{+ 0.21 }_{ -0.10}$ & $> 4.59$ & {$ 4.40 ^{+ 0.10 }_{ -0.11 }$}  \\
        ~ & vturb ($10^{3}$~km s$^{-1}$) & $ 0.3 ^{+ 1.4 }_{ -0.2} $ & $ > 7.2 $ & $ < 6.2 $ & {$ < 0.98 $} & {$ < 0.18$}  \\
        ~ & z & $ -0.068 ^{+ 0.002 }_{ -0.001 } $& $ -0.052 ^{+ 0.002 }_{ -0.001 }$ & $ -0.066 ^{+ 0.002 }_{ -0.003 } $ & {$ -0.146 ^{+ 0.020 }_{ -0.002 }$} & {$ -0.088 ^{+ 0.013 }_{ -0.011 }$}  \\\hline
        {\tt diskbb} & normalization ($10^{3}$) & $ 10.07 ^{+ 0.04 }_{ -0.06 } $ & $ 7.70^{+ 0.04 }_{ -0.05 }$ & $ 8.05 ^{+ 0.08 }_{ -0.10 }$ & {$ 6.20 ^{+ 0.11 }_{ -0.12 }$} & {$ < 0.49 $}  \\\hline
        {\tt powerlaw} & normalization ($10^{-2}$ photons s$^{-1}$ cm$^{-2}$ keV$^{-1}$) & $ 1.264 ^{+ 0.003 }_{ -0.003 }$&  $ 1.064 ^{+ 0.002 }_{ -0.003 } $ & $ 1.04 ^{+ 0.05 }_{ -0.05}$ & $ 1.068 ^{+ 0.006 }_{ -0.006 } $
 & {$ 1.284 ^{+ 0.007 }_{ -0.010 }$}  \\\hline
        \multirow{3}{*}{${\tt G_\text{Fe}}$} & Energy (keV)& $ 6.64 ^{+ 0.03 }_{ -0.04 } $
& $ 6.56 ^{+ 0.08 }_{ -0.08 } $ & $ > 6.82 $ & $ 6.75 ^{+ 0.10 }_{ -0.09 } $& {-}  \\
        ~ & $\sigma~(\rm{keV})$ & $< 0.14$& $ < 0.21$
 & $ < 0.48$ & $0.001 ^{*1}$ & {-}   \\
        ~ & normalization ($10^{-5}$ photons s$^{-1}$ cm$^{-2}$) & $ 0.53 ^{+ 0.19 }_{ -0.19 } $ & $ 0.58 ^{+ 0.23 }_{ -0.23 } $ & $  0.41 ^{+ 0.33 }_{ -0.35 } $ &  $ 0.98 ^{+ 0.43}_{ -0.43 } $
 & {-}   \\\hline
        \multirow{4}{*}{${\tt const_\text{cross}}$} & FPMA/XMM-pn & -& - & - &  1.28(fixed) & {-}   \\
        ~ & FPMB/XMM-pn & -& - & - & 1.24(fixed) & {-}   \\
        ~ & FPMA/Swift-XRT & -& - & - & {-}  &  1.25(fixed)  \\
        ~ & FPMB/Swift-XRT & -& - & - & {-}  &  1.21(fixed)  \\\hline
        \hline
        \multicolumn{2}{c}{$\chi ^2$/d.o.f} & 1.13 & 1.13&1.01& 1.00 &{1.47}\\\hline
        \end{tabular}
\begin{tablenotes}
\item[*1] not constrained
\end{tablenotes}
\end{threeparttable}
\end{table*}

\section{discussion}
\subsection{Geometry of the disc and torus}
In the spectral analysis,
we introduced an outer disk/torus reflection represented by the {\tt pexmon} model, where
 the solid angle  of the outer disk or torus
 was $\Omega/2 \pi = 0.20\pm0.01$ (with the triple partial covering model). 
On  one hand, 
if there is an isotropic radiation source  on nearly  infinite size of the 
reflector, the solid angle of the reflector  will be as large as $\Omega/2\pi  \approx 1$. 
On the other hand, solid angle of the upper-half of the outer torus with a half-opening angle $\theta$, seen from the
central illuminating source, is estimated as  $\Omega/2\pi \approx   \cos{\theta}$.
Therefore,
the half-opening angle of the torus is estimated to be $78\pm1$ deg,
equivalent to the torus angular width $90 - (78\pm3) = 12\pm1$ deg.  
 Recently, it has been pointed out that  Seyfert galaxies commonly have  X-ray clumpy torii
with small
angular widths, 
such as $\sim$15 deg in the Circinus galaxy \citep{tani19}, $\sim$10 deg in IC 4329A \citep{oga19}, and $\sim$20 deg in NGC 7469 \citep{oga19}.
The small solid angle of the cold reflector we found in Mrk 766 is consistent with the reflection 
from such an  outer clumpy torus with relatively small angular width.

Also, we estimate  the inner disk reflection rate of the {\tt relxill} model. 
We  compare  ${\tt const_\text{ref}}*{\tt relxill}$ and
{\tt rdblur}*{\tt pexrav} \citep{rdblur,mag95}, because the latter gives a more intuitive 
interpretation of the normalization. 
 In this manner,
we estimate the inner disk reflection rate to be 
$\Omega / 2\pi \sim 0.3$.
We assume that the X-ray radiating  corona is isotropic
and slightly extended, so that it is 
{\em  partially} covered by smaller clumps in the outer parts.
It is difficult to estimate the corona size, but let's assume 
a radius of $\sim$10 $R_g$.
The solid angle of the inner reflecting disk, seen from a point above the black hole,  is estimated as $\Omega/2\pi \approx  \cos{\phi} - \cos{\psi}$, where  $\phi$ is the angle between the black hole and the innermost radius, and $\psi$ is the one between the black hole and the outermost radius.
If seen from the height of 10 $R_g$ above the black hole, 
 the inner disk  reflection with  $r_\text{in} =  6 \: R_g$ and  $r_\text{out} = 400 \: R_g$ gives 
 the  solid angle as large as  $\Omega/2\pi \approx 0.8$, that 
 is much greater than the observed value. Thus,
 it is more likely that the  inner edge of the disk does {\em not}\/ reach the ISCO (6 $R_g$), but is rather 
{\em  truncated}\/ at around a few tens of $R_g$.
If the X-ray corona size  is $\sim$10 $R_g$, the innermost radius  of the disk is 
$\approx$30 $R_g$ 
to give 
the observed value $\Omega / 2\pi \approx 0.3$.

In the spectral fitting with the {\tt relxill} model, 
we fixed the inner disk radius ($r_\text{in} = 6 \: R_g$) and the black hole spin ($a = 0$) assuming a  Schwarzschild black hole. 
Even when we changed the inner disk radius in the range of  6 $R_g$ $\le$ $r_\text{in}$ $\le$ 50 $R_g$ with $a=0$ 
and all the other parameters fixed, the spectral fitting did not show significant changes.
Similarly, even when we changed the black hole spin (0 $\le$ $a$ < 1) with the disk inner radius fixed to the ISCO, the spectral fitting result hardly changes.
Therefore, we can hardly constrain the inner disk radius and the black hole spin in our model.

\subsection{Geometry of partial covering clouds and UFOs}

\begin{figure*}
\centering
    \includegraphics[width=15cm, clip]{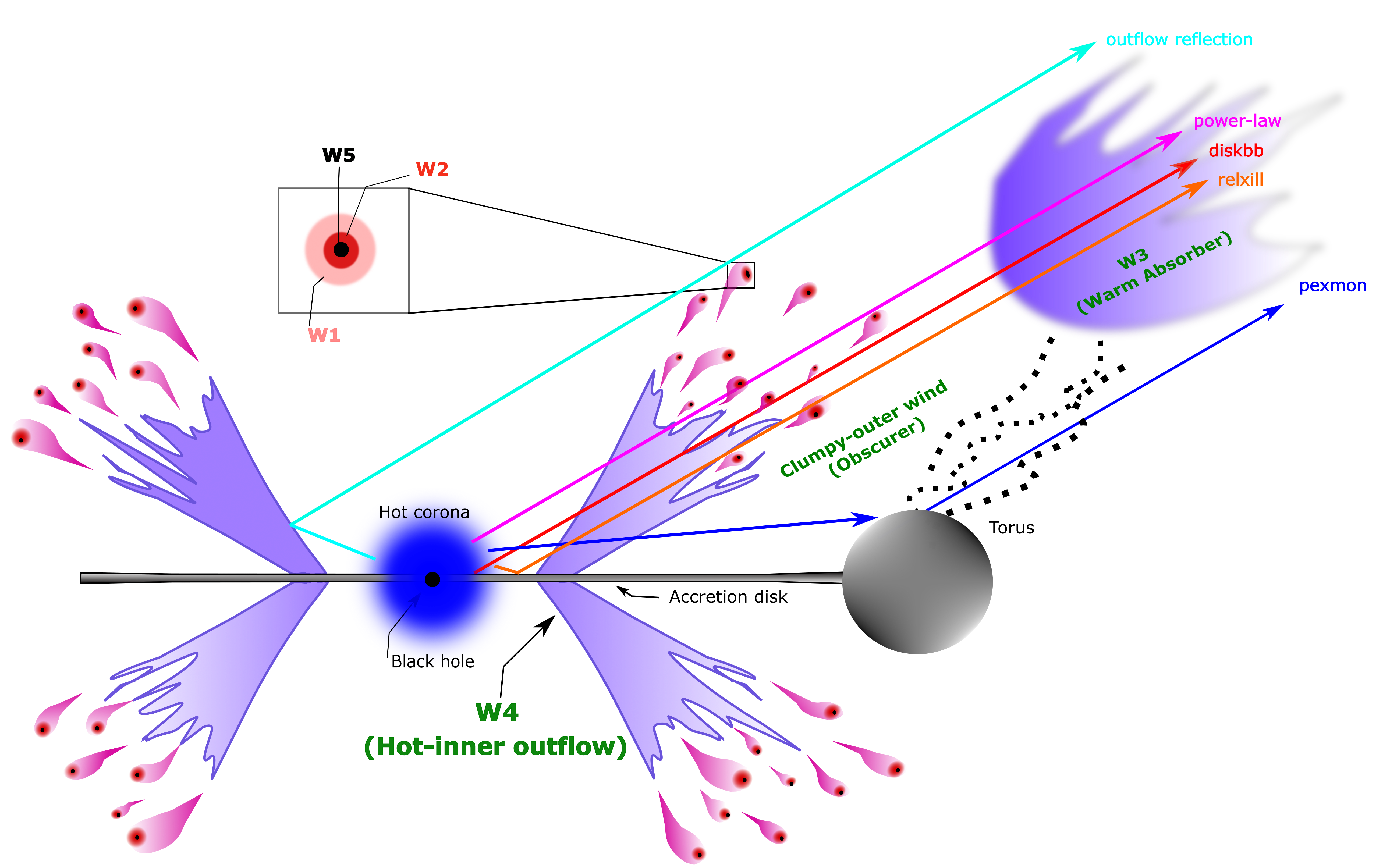}
    \caption{Assumed geometry of Mrk 766. 
    Arrows with different colors represent different X-ray emission components. 
    There are three distinct intervening absorbers; the clumpy-outer winds with internal layers (W$_1$, W$_2$, and W$_5$),
    the hot-inner outflow corresponding to the UFO (W$_4$), and the warm absorber (W$_3$).  
    The original figure made for NGC5548 by \citet{mido21}  was modified considering the differences between the
    two sources.
     }
    \label{fig:geometry}
\end{figure*}

The schematic picture of the clumps in the wind is shown in Fig.\ \ref{fig:geometry}.
For the geometry of the partial covering clouds,
\citet{ris11} proposed that the absorber in Mrk 766 has a cometary-like structure with a neutral core and an ionized tail.
Also, the double-layer absorber with internal structure has been proposed in other AGNs (e.g., \citealt{miy12,iso16}).
Another scenario suggests that the cold clumps are embedded in the ionized wind and
are subjected to the  X-ray illumination and a pressure balance \citep{don16,elv00}.
This will invoke a thermal instability in the ionized gas and the emergence of cooler clumps \citep{kro81}.
Furthermore, recent numerical hydrodynamic simulations predict  the emergence of  clumpy outflows 
\citep[e.g.,][]{Takeuchi13, Kobayashi18}, 
so that  the ``hot-inner and clumpy-outer winds" model successfully explains spectral variations 
of several bright Seyfert galaxies 
\citep{Mizumoto19, mido21}. 

\begin{figure}
\centering
	\includegraphics[width=8cm]{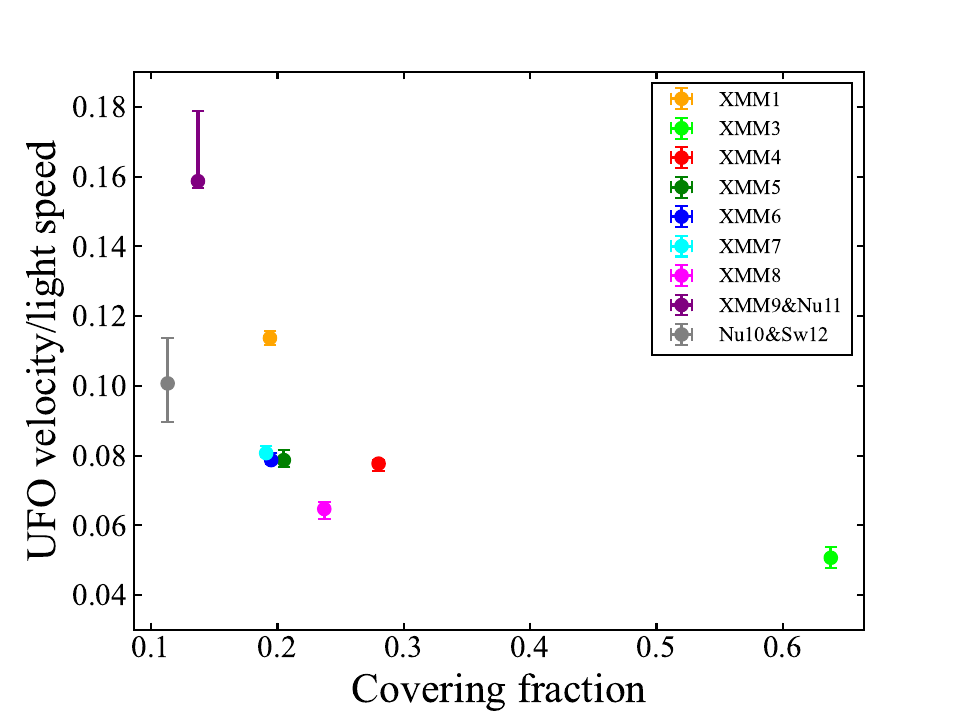}
    \caption{Relation between the partial covering fractions 
    (with the triple partial covering model) and the UFO velocities. XMM2 data are not used where the UFO feature 
    was not clear. 
    }
    \label{fig:cf_velo}
\end{figure}

Regardless of minor differences in  the models and interpretations, it is generally accepted 
that there are fast and hot flows near the black hole (i.e., UFO), and the partial covering clouds   farther 
from the black hole. 
We find that the partial covering fraction of the  partial covering clouds is lower when the UFOs velocity gets higher (Fig. \ref{fig:cf_velo}).
This relation is exactly consistent with the results of hydrodynamic simulations that  
the  covering fraction and the UFO velocity anti-correlate as long as the  mass outflow rate and the clump formation radius are invariable 
(see, e.g.,  Eq.\ 19 in  \citealt{Takeuchi13}).
We note that 
this correlation is statistically significant at the 90 \% confidence level but not 95 \%, with the Pearson coefficient of $-0.61$ and the p-value of 0.08.
It will be important to determine more accurately the UFO velocities and partial covering fractions in future microcalorimeter observations.

The UFO location  is estimated as follows;
the minimum radius is estimated by the balancing equation between gravitational potential and kinetic energy, as 
\begin{equation}
     r_{\mathrm{min}} = \frac{2 G M_{\mathrm{BH}}}{v_{\mathrm{out}}^2}, 
     \label{eq:miniufor}
\end{equation}
and the maximum radius is estimated from the definition of the ionization parameter under the condition that $nr \sim$ $N_{\rm H}$ and $L$ is the X-ray luminosity in 13.6~eV--13.6~keV, as
\begin{equation}
r_{\mathrm{max}}= \frac{L}{\xi N_{\rm H}}.
\label{eq:maxufor}
\end{equation}
The average parameters of UFOs in this study are $N_{\rm H} \sim 2.1 \times 10^{23}$cm$^{-2}$, $\log \xi \sim 4.2$ and velocity$\sim -0.10 c$.
Then, the UFO is considered to locate in the range of  $\sim10^2\,R_g - 10^3\,R_g$.

\subsection{Spectral variability}
The right panel of Fig.\  \ref{fig:Cf-norm} depicts the power-law normalization and the partial covering fraction in the triple partial covering model.
The power-law normalization exhibits a change of a factor of $\sim$1.5 while  the covering fraction  changes by a factor of $\sim$5.
This implies that while the hard X-ray variation ($\gtrsim 10$ keV) is due to the  power-law normalization changes,  the soft X-rays ($\lesssim 10$ keV) are more significantly affected  by   the partial covering fraction.
We have  explained the spectral variability  over the  $\sim$15 years assuming no
changes of the power-law photon index,  the column densities and ionization parameters
of the tripel-layered partial absorbers (W$_1$, W$_2$ and W$_5$)
and the  warm absorber (W$_3$).

Our spectral datasets consist of the 10 observations, each of which has
$\sim$day exposure, over $\sim$15 years (Table \ref{tab:xmm}).
All the UFO parameters (W$_4$), namely the column density, ionization state and velocity, 
are variable for different spectra,  
as well as  the partial covering fraction.
\cite{risa-else11} examined  timescales  of the partial covering change
 from the flux and hardness-ratio variations. 
They estimated that the location of the partial covering clouds is at $\sim 10^3\,R_g - 10^4\,R_g$.
We naturally think that the partial covering clouds are outer UFO with respect to these locations.
Considering that the corona (a few tens of $R_g$) is {\em  partially} covered by the partial covering clouds, 
we estimate the size of the partial covering clouds $\lesssim 10^{13}$ cm.  

These differences in the  variation timescales of the different  absorbers reflect  locations of the multiple  absorbers around the black hole,
and are consistent with the ``hot-inner and clumpy-outer wind scenario"
\citep[Fig.\ \ref{fig:geometry}; ][]{Mizumoto19}. The hot-inner outflow (UFO) is produced at the inner-part of the accretion disk, which  turns into the clumpy-outer wind (partial covering clouds, a.k.a.\ ``obscurers"; \citealt{2014Sci...345...64K}). 
The warm absorber may be produced by irradiation and evaporation of
the outer torus \citep{miz19}.

\subsection{Scattering in the ionized absorber} \label{sec:Gorigin}
We have introduced a positive Gaussian at 6.4--6.7~keV as ${\tt G_{Fe}}$ (\S\ref{sec3.3}; Fig.\ \ref{fig:nogauss}).
 Importance of the scattering by surrounding material in Mrk 766  has already been pointed out \citep[e.g., ][]{mil-and-tur06, Tur07, sim08},
so that the  origin of the Gaussian line is likely  the Fe-K resonance scattering  produced by the ionized material out of the line-of-sight. 
The average peak energy of the emission line is $\sim$6.61~keV, which is consistent with the ionization state of the UFO ($\log \xi \sim 4.2$) (e.g., \citealt{kal01,kal04}), 
supporting  the idea that this line results from the resonance scattering in the UFO. 
In fact, 
\cite{mizu21} suggested  that the UFO material out of the line-of-sight 
should induce resonance scattering
line (shown with the cyan line in Fig. \ref{fig:geometry}).
Via X-ray spectral synthesis with a radiation hydrodynamic simulation, 
\cite{mizu21} demonstrated that the equivalent width (EW) of the Fe-K resonance emission lines ranges from $\sim$$0.02-0.05$~keV and remains relatively constant with respect to the viewing angle, while
that of the UFO's absorption lines experiences a significant change from $-0.18$~keV to null  as the viewing angle changes \citep[see also, e.g.,][]{hagi15}. 
    The observed EWs of the ${\tt G_{Fe}}$ and the UFO's absorption line (Table \ref{tab:ew})  are consistent with the simulation when the 
UFO is observed at  an inclination angle of 45$^\circ$ -- 55$^\circ$.

\begin{table}
    \centering
    \caption{The emission/absorption equivalent width of Fe-K with the triple partial covering model. 
    }
    \begin{threeparttable}
    \begin{tabular}{|c|c|c|c|c|}
    \hline
        Data & Emission EW (keV) & Absorption EW (keV)$^{*1}$ \\ \hline
        XMM1 & 0.063$^{+0.029}_{-0.029}$ & $-0.134$ \\ 
        XMM2  & 0.187$^{+0.019}_{-0.019}$ & $-0.001$ \\ 
        XMM3 & 0.056$^{+0.018}_{-0.017}$ & $-0.164$ \\ 
        XMM4  & 0.056$^{+0.017}_{-0.017}$ & $-0.122$ \\ 
        XMM5 & 0.061$^{+0.032}_{-0.032}$ & $-0.073$ \\ 
        XMM6 & 0.031$^{+0.011}_{-0.011}$ & $-0.068$ \\ 
        XMM7 & 0.039$^{+0.015}_{-0.015}$ & $-0.092$ \\ 
        XMM8 & 0.037$^{+0.030}_{-0.032}$ & $-0.063$ \\ 
        XMM9 \& Nu11  & 0.067$^{+0.029}_{-0.029}$ & $-0.108$ \\
        Nu10 \& Sw12 & 0 & $-0.216$ \\ \hline
        Average  & 0.052 & $-0.104$ \\ \hline
    \end{tabular}
    \label{tab:ew}
\begin{tablenotes}
\item[*1] Errors are less than  1\%.
\end{tablenotes}
\end{threeparttable}
\end{table}

\section{Conclusion}

In this paper, we tried to solve the origins of the complex iron K-line structure and the significant X-ray spectral variations in 0.3 -- 79 keV of Mrk 766 with a simple model.
We used 10 spectral datasets taken over $\sim$15 years from 2000 to 2015; nine observation sequences of  \textit{XMM-Newton}, one of which was simultaneous with \textit{NuSTAR}, and 
another observation of \textit{NuSTAR} which was 
simultaneous with  \textit{Swift}.
The energy spectra are largely  
explained by a  simple geometry that a moderately extended central  
X-ray region emitting a power-law spectrum is affected by
several distinct absorption zones in the line-of-sight; 
the hot and fast inner outflow (UFO), the outer-clumpy winds (partial covering clouds) comprising multiple ionization layers, and the warm absorber further outward.  
Spectral variations are mainly explained due to changes in the partial covering fraction, the power-law
normalization, and the UFO parameters. 
Other spectral parameters are not significantly
variable over $\sim$15 years.

Additionally, a slightly broadened Fe-K emission line is  required  at $\sim$6.4--6.7keV with
the intrinsic  width (1 $\sigma$) 0.1--0.3~keV and the equivalent width  
$\sim$0.05~keV.
This emission line is considered to be produced by the X-ray scattering in the UFO out of the line-of-sight.

\section*{Acknowledgements}
We thank the anonymous referee for valuable comments and suggestions to improve this paper. 
This research has made use of  data and/or software provided by the High Energy Astrophysics Science Archive Research Center (HEASARC), which is a service of the Astrophysics Science Division at NASA/GSFC.
We used  data by \textit{XMM-Newton} led by the European Space Agency's (ESA).
This study also used  data of \textit{NuSTAR} led by the California Institute of Technology (Caltech) and managed by NASA's Jet Propulsion Laboratory in Pasadena, and of \textit{Swift} which is based on the Mission Operations Center (MOC) operated by 
the Pennsylvania State University and NASA.
The authors used JAXA Supercomputer System generation 3 (JSS3).
This research was supported by JSPS KAKENHI Grant Number JP21K13958 (MM). MM acknowledges support from the Hakubi project at Kyoto University.
The authors thank Prof.\ Kyoko Matsushita and Dr.\ Shogo B. Kobayashi for discussion.
The authors thank Ms.\ Juriko Ebisawa for the artwork  (Fig.\ \ref{fig:geometry}).

\section*{DATA AVAILABILITY}
Al the observational data used in this paper are publicly available.
They can be accessed via  HEASARC Browse:\\
\url{https://heasarc.gsfc.nasa.gov/db-perl/W3Browse/w3browse.pl}

\section*{ORCID iDs}
Yuto Mochizuki~${\orcidlink{0000-0003-3224-1821}}$\\
\url{https://orcid.org/0000-0003-3224-1821}\\
Misaki Mizumoto~${\orcidlink{0000-0003-2161-0361}}$\\
\url{https://orcid.org/0000-0003-2161-0361}\\
Ken Ebisawa~${\orcidlink{0000-0002-5352-7178}}$\\
\url{https://orcid.org/0000-0002-5352-7178}

%%%%%%%%%%%%%%%%%%%%%%%%%%%%%%%%%%%%%%%%%%%%%%%%%%



%\bibliographystyle{mnras}

\bibliographystyle{mnras}
\bibliography{00}


%%%%%%%%%%%%%%%%%%%%%%%%%%%%%%%%%%%%%

%%%%%%%%%%%%%%%%% APPENDICES %%%%%%%%%%%%%%%%%%%%%

%%%%%%%%%%%%%%%%%%%%%%%%%%%%%%%%%%%%%%%%%%%%%%%%%%

% Don't change these lines
\bsp	% typesetting comment
\label{lastpage}
\end{document}